# Production, Characteristics and Biological effects of Protonated Small Water Clusters

Yixin Zhu

Hangzhou Shanshangshui Technology Co., LTD, Hangzhou, China

**Abstract**

**Significance:** The production and characteristics of protonated small water clusters (**PSWCs**) were reported in this work, where in electrospray ionization (ESI) of pure water, the species obtained were singly charged molecular ions consisting of 2, 3, 4 or 5 water molecules attached to a hydrogen ion, $[(H_2O)_n+H]^+$, where n = 2, 3, 4 or 5. We proposed a new type of **PSWCs** structure: 2, 3, 4, 5 water molecules wrapped around a hydrogen ion which is located at the electrical and geometric center, forming a very stable molecular structure. Furthermore, biological tests of the **PSWCs** on mitochondrial function of intestinal epithelial cells and liver cells in mice showed the better therapeutic effect on inflammatory bowel diseases compared to that of the biologic agent Infliximab.

**Aim:** To produce stable PSWCs as a new source of nutrient for health.

**Approach:** The **PSWCs** were produced by electrospray technology, and identified by an electrospray mass spectrometer. The biological effects of the **PSWCs** on mitochondrial function of intestinal epithelial cells and liver cells in mice are carefully checked.

**Results:** The production, characteristics and the biological effects of the **PSWCs** are studied, that we first generate the **PSWCs** by electrospray mechanism, which is proved to be charged (with positive charge). The obtained **PSWCs** are composed of 2, 3, 4, 5 water molecules plus a proton, respectively. The test of biological effects shows that the **PSWCs** can reduce radicals and improve the cell functions, indicating very important biological functions of the PSWCs.

**Conclusions:** **PSWCs** was successfully produced by electrospray technology in our lab, the produced PSWCs were very stable under normal conditions without any obvious concentration changes of hydrogen ions in the last 3 years after produced, even the water is heated to 100 degrees Celsius. Furthermore, the effect of PSWCs on mitochondrial function of intestinal epithelial cells and liver cells in mice showed that the therapeutic effect of PSWCs on inflammatory bowel disease was better than that of the biologic agent Infliximab, where the higher the concentration of hydrogen ions, the better the therapeutic effect is.

Author, E-mail: zyx@hprotonwater.com

## 1 Introduction

The importance of protons in life is self-evident, all living things are composed of cells, almost all the energy of eukaryotic cells is provided by mitochondria, and the raw material of the mitochondria to produce energy molecule ATP is proton.[1] Therefore, it can be considered that proton is the indispensable for life. The production of hydrogen ions is very simple and can be produced by electrolysis. However, the hydrogen ions produced by this method directly contact with the cathode of the electrolytic cell, leading to the formation of hydrogen and volatilize, and therefore, no stable excess proton exists in water.

Herein, to produce the PSWCs, we use a special method to generate hydrogen ions (protons) under the condition of high electric field and realize the chelation of protons and water molecules in the high electric field area. A proton wrapped by 2, 3, 4, 5 water molecules, respectively to form a hydrogen ion charged small molecular group of water was then observed, such kind of hydronium ion charged small molecular group of water is very stable permanently.

The generation process of PSWCs is as follows: pure electrically neutral water is sent to the hydrogen ion generator, and the strong electric field breaks the raw water into clusters of water molecules, some of which consist of 1, 2, 3, 4, and 5 water molecules. Due to the action of high electric field, a single water molecule is broken into a proton ($H^+$) and a hydroxide ($OH^-$), and the hydroxide moves towards and eventually contacts the anode, transporting electrons to the anode. Each of the four hydroxide group loses an electron, forming two parts of water ($H_2O$) and one part of oxygen ($O_2$). At the same time, the proton moves away from the anode, whereas the oxygen atomic ends of the water in the small molecular cluster composed of 2, 3, 4 and 5 water molecules are convex toward the anode. Under the action of high electric field, the proton and the oxygen atomic end of the small molecular cluster water attract each other, forming a tightly bound structure with electrostatic force. The stable $H^+(H_2O)_n$ molecular clusters are formed while drifting further into the field free region, where n = 2, 3, 4 or 5 .

Furthermore, in this work, we also perform biological function test of the PSWCs on mitochondrial function of intestinal epithelial cells and liver cells in mice. The test results showed that the PSWCs have the better therapeutic effect on inflammatory bowel diseases compared to that of the biologic agent Infliximab.

## 2 Theory

As shown in Fig.1, a capillary (1) with an inner diameter of 10-100 nm is fed with pure water, and a DC or pulsed positive voltage power supply (3) is added to the capillary forming a strong electric field on the tip surface. When the electric field force is greater than the indicated tension of the liquid, droplets will leave the Taylor Cone (3) surface. The voltage is the initial spray voltage given by

$$V_{on} \approx 2\times 10^5 (\gamma R_c)^{\frac{1}{2}} \ln \left(\frac{4d}{R_c}\right), \tag{1}$$

where γ is surface tension of the solvent, $R_c$ is the outer radius of the spray capillary and d is the distance between the spray tip and the counter-electrode. In our setup, γ = 0.073 N/m for $H_2O$, $R_c$ = 12.5 $\mu m$ and d = 1.0 mm, therefore Von = 1.1 kV, which matches well with our experimental value.

The electric field at the tip of the Taylor cone is:

$$E_c \approx \frac{2V_c}{r_c \ln(4d/r_c)},  \qquad (2)$$

Here $r_c$ is the radius at the tip of the Taylor cone, not the $R_c$ of capillary. We estimate $r_c$ is on the order of 100 to 600 nm. Therefore, $E_c$ = 2.1 × 10$^9$ V/m to 4.16 × 10$^8$ V/m. And the electric field energy density is:

$$W_e = \frac{1}{2} \varepsilon_0 E^2, \qquad (3)$$

Here $\varepsilon_0$ = 8.85 × 10$^{-12}$ C$^2$/Nm$^2$. Therefore $W_e$ = 1.9 × 107 J/m$^3$ to 7.7 × 105 J/m$^3$. On the other hand, if we assume the gas phase $H_2O$ is ideal gas, we have:

$$PV = nRT, \qquad (4)$$

Where P is the vapor pressure, V is the volume of gas, n is the mole number, R is ideal gas constant (8.314 Pa m$^3$ mol$^{-1}$ K$^{-1}$) and T is temperature. At room temperature of 298k (25 °C), vapor pressure of $H_2O$ is 3173 Pa. Therefore, its concentration can be written as:

$$\frac{n}{V} = \frac{P}{RT} = 1.28 \ mol/m^3, \qquad (5)$$

The O-H bond energy is 464 kJ/mol or 4.64 × 10$^5$ J/mol. So the energy density to break the O-H bond must be larger than

$$4.64 \times 10^5 \frac{J}{mol} \times 1.28 \frac{mol}{m^3} = 5.94 \times 10^5 \ J/m^3 \qquad (6)$$

Therefore, the electric field energy $W_e$ is more than enough to break the O-H bond and produce H$^+$.

When the electric field energy density at the tip is greater than the hydrogen bond energy (18.84 kJ/mol) of water, the raw water clusters are broken into small water clusters, such as $(H_2O)_n$ (n= 1, 2, 3, 4, 5 ..., respectively) . When the energy density of the electric field at the tip is greater than the (O-H) bond energy (464 J/mol) of water, a single water molecule in the gas phase will be heterosplit into a proton and a hydroxide separately, where 4 hydroxides will be absorbed by the anode, leading to the formation of one oxygen and two water molecules as:

$$4OH^- - 4e = 2H_2O + O_2 \qquad (7)$$

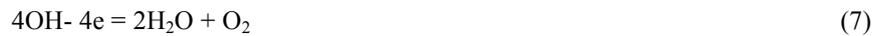

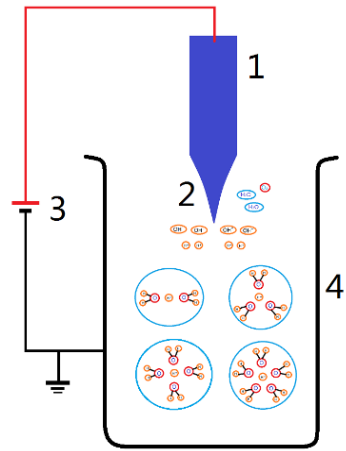

**Fig. 1** Schematic diagram of (**PSWCs**).

At the same time, clusters containing 2, 3, 4 and 5 water molecules near the Taylor Cone are polarized. [2] The oxygen atom end of the water molecular cluster is convex towards the anode, and the proton just leaving the anode is absorbed. Due to the action of strong electric field, they are tightly adsorbed together and drift toward the cathode. In the drift process of the field-free zone, a stable hydrogen ions charged small molecular water clusters was formed by 2, 3, 4, 5 water molecules wrapped around a proton. Due to the insulation effect of the surrounding water molecules, hydrogen ions cannot contact the cathode, thus ensuring the stable existence of hydrogen ions in water.

## 3　Method

### 3.1　Production of *PSWCs*

**PSWCs** were produced from an electrospray equipment based on the new charging mechanism, and used in validation and functional tests. To test the hydrogen ion stability of the PSWCs, it was placed in a grounded metal container, heated to boiling, and allowed to cool down, and the hydrogen ion concentration was measured and the hydrogen ion concentration is the same as before heating.

### 3.2.1 DSS-induced Colitis

To establish DSS-induced colitis models, mice were treated with DSS (2% w/vol; molecular weight: 36 to 50 kDa; #160110, MP Biomedicals) in drinking water for 6 days, followed by regular drinking water. At the end of the treatment, the mice were sacrificed, and colonic tissues were collected for further examination.

Disease activity index (DAI) scores were used to evaluate clinical signs of mouse colitis, including body weight loss, occult blood, and stool consistency.[4] Mice were scored blindly for the colitis phenotype. The weight loss score was graded as follows: 0, no weight loss; 1, loss of 1-5% original weight; 2, loss of 6-10% original weight; 3, loss of 11-20% original weight; 4, loss of >20% original weight. The bleeding score was determined as follows: 0, no blood as determined by Hemoccult (Beckman Coulter) analysis; 1, positive Hemoccult; 2, visible blood traces in stool; 3, gross rectal bleeding. The stool score was determined as follows: 0, well-formed pellets; 1, semi-formed stools that did not adhere to the anus; 2, pasty semi-formed stool that adhered to the anus; 3, liquid stools that adhered to the anus.

*3.2.2 Hematoxylin and Eosin (H&E)*

The mouse intestinal tract and liver were obtained immediately after sacrifice, opened longitudinally, washed with 70% ethanol. It was then fixed in a 10% formalin solution overnight, embedded in paraffin, and sliced into sections. Deparaffinization was performed by incubating the sections in xylene for 30 min, rehydration in decreasing ethanol solutions (100%, 95%, and 80%) for 5 min each, and then distilled water for 5 min. H&E staining was performed by immersing the sections in hematoxylin for 5 min, followed by PBS for 3 min to prevent background staining, then eosin for 2 min, and finally distilled water for 5 min. The stained sections were mounted using neutral resin and a coverslip and observed and digitally photographed using a Leica DM2000 LED microscope.

*3.2.3 Oil Red Staining*

The mouse liver sections was incubated in xylene for 30 min, rehydration in decreasing ethanol solutions (100%, 95%, and 80%) for 5 min each, and then distilled water for 5 min. Completely cover the tissue sections with the Oil Red working solution. Incubate the slides in a horizontal position at room temperature for 20 minutes. Followed by PBS for 3 min to prevent background staining, then eosin for 3 min, and finally distilled water for 10 min. The resulted observed and digitally photographed using a Leica DM2000 LED microscope.

*3.2.4 Histological Analysis*

A board-certified pathologist performed histological assessment of colitis blindly based on previously described criteria, including the extent and severity of inflammation and ulceration of the mucosa. The severity score for inflammation was as follows: 0, normal (within normal limit); 1, mild (small, focal, or widely separated, limited to the lamina propria); 2, moderate (multifocal or locally extensive, extending to the submucosa); 3, severe (transmural inflammation with ulcers covering >20 crypts).

*3.2.5 Immunoblot*

The RIPA buffer (#89900, Thermo Fisher Scientific) supplemented with a protease inhibitor cocktail (#11697498001, Roche) was used to extract proteins from colon, mucus layer, and stool samples. The protein concentration was measured using a BCA protein assay kit (#23225, Thermo Fisher Scientific). The resolved proteins were then subjected to 15% sodium dodecyl sulfate-polyacrylamide gel electrophoresis (SDS-PAGE). the separated proteins were transferred onto a 0.45 μm polyvinylidene fluoride (PVDF) membrane (#88518, Thermo Fisher Scientific), which was then blocked with 5% BSA in Tris-buffered saline containing 0.05% Tween 20 (TBST) for 1 h. The membrane was washed in TBST for three times and then incubated with primary antibodies Total OXPHOS Rodent (#ab110413, Abcam) or mouse anti-ACTB antibody (1:2000, #60008-1-Ig, Proteintech) diluted in 5% BSA on a rocker overnight at 4°C. After washing with TBST for another three times, the membrane was incubated with HRP-conjugated secondary antibody (#31430 or #31460, Thermo Fisher Scientific) diluted in 5% BSA for 1 h at room temperature. The chemiluminescence signals were detected using SuperSignal West Pico Chemiluminescent Substrate (#34580, Thermo Fisher Scientific), and the images were captured using GE Healthcare Life Sciences Amersham Imager 600.

*3.2.6 Real-time qPCR Analysis*

Total RNAs from colonic tissue and liver tissue of DSS-treated mice were purified via precipitation with lithium chloride as reported previously. The mRNA level was evaluated using a reverse transcription reaction with Moloney Murine Leukemia Virus (M-MLV) reverse transcriptase (#639574, Takara), followed by qPCR analysis with SYBR Premix Ex Taq (#RR420A, Takara) on a Roche 480 real-time PCR system. The primer sets used for qPCR are listed in Table 1.

*3.2.7 Oxidative stress reactive oxygen species (ROS) detection*

The unfixed frozen slices of intestinal and liver from mice were first washed by cleaning liquid for 5min, and then add ROS dye probe working solution (#HL10016.6 Shanghai haling biological technology) incubated for 60min without light at 37 °C. After washing slices with PBS 3 times, nuclei were stained with Hoechst 33342 (10 μg/mL, #H3570,

Thermo Fisher Scientific) for 30 min at room temperature. Fluorescence images were acquired at randomly selected locations using a Nikon A1 confocal microscope system.

Table 1 RT-qPCR Primers.

| Gene | species | Forward primer (5′→3′) | Reverse primer (5′→3′) |
|---|---|---|---|
| *Actb* | mouse | GTGACGTTGACATCCGTAAAGA | GCCGGACTCATCGTACTCC |
| *Il-1β* | mouse | GAAATGCCACCTTTTGACAGTG | TGGATGCTCTCATCAGGACAG |
| *Il-6* | mouse | TCTATACCACTTCACAAGTCGGA | GAATTGCCATTGCACAACTCTTT |
| *Il-17a* | mouse | GGCCCTCAGACTACCTCAAC | TCTCGACCCTGAAAGTGAAGG |
| *S100a8* | mouse | AAATCACCATGCCCTCTACAAG | CCCACTTTTATCACCATCGCAA |
| *Tnfα* | mouse | CTGAACTTCGGGGTGATCGG | GGCTTGTCACTCGAATTTTGAGA |
| *Cxcl1* | mouse | ACTGCACCCAAACCGAAGTC | TGGGGACACCTTTTAGCATCTT |
| *Cxcl2* | mouse | CCAACCACCAGGCTACAGG | GCGTCACACTCAAGCTCTG |
| *Ccl2* | mouse | AACTCTCACTGAAGCCAGCTCT | CGTTAACTGCATCTGGCTGA |
| *Ccl3* | mouse | TGTACCATGACACTCTGCAAC | CAACGATGAATTGGCGTGGAA |
| *Gcsf* | mouse | ATGGCTCAACTTTCTGCCCAG | CTGACAGTGACCAGGGGAAC |
| *Pgc1α* | mouse | CAGCCT CTTTGCCCAGATCTT | TCACTGCACCACTTGAGTCCAC |
| *TFAM* | mouse | GGAATGTGGAGCGTGCTAAAA | ACAAGACTGATAGACGAGGGG |
| *CS* | mouse | GGACAATTTTCCAACCAATCTGC | TCGGTTCATTCCCTCTGCATA |
| *Aco2* | mouse | ATCGAGCGGGGAAAGACATAC | TGATGGTACAGCCACCTTAGG |
| *Idh3a* | mouse | CCTCCTGCTTAGTGCTGTGA | CGTTGCCTCCCAGATCTTT |
| *Ndufa4* | mouse | TCCCAGCTTGATTCCTCTCTT | GGGTTGTTCTTTCTGTCCCAG |
| *Shda* | mouse | GGAACACTCCAAAAACAGACCT | CCACCACTGGGTATTGAGTAGAA |
| *Cyc1* | mouse | GCTCCTCCCATCTACACAGAAG | ATGGTCATGCTCTGGTTCTGA |
| *Cox5a* | mouse | GGGTCACACGAGACAGATGA | GGAACCAGATCATAGCCAACA |
| *Atp5g1* | mouse | AGTTGGTGTGGCTGGATCA | GCTGCTTGAGAGATGGGTTC |

*3.2.8 Assessment of mitochondrial membrane potential (Δψm) using JC-1*

JC-1 is a lipophilic, cationic dye that can selectively enter into mitochondria and reversibly change color from green to red as the membrane potential increases. For spectrofluorimetric measurement ($1 \times 10^6$ cells/mL cell suspension buffer) were incubated with 1 μL (5 μM) JC-1 for 25 min in dark with gentle agitation. Cells were then washed twice with fresh media, resuspended in a total volume of 2 mL and transferred into the fluorimeter cuvette. The peak excitation of 560 nm with emission filter being centered at 595 nm (25 nm bandwidth) was used for detecting the red fluorescence (J-aggregates). The green fluorescence (JC-1 monomers) was detected with excitation and emission at 485 nm and 535 nm respectively. The results were expressed in terms of arbitrary fluorescence units. Similarly confocal microscopic assessment of Δψm involved incubation with JC-1 (5 μM) in dark for 25 min as described above and the cells were washed with fresh media. About 50 μL of the cells was seeded onto slides and was spread gently. The cells were then covered with coverslip using xylene and DPX Mountant and analyzed with Nikon A1 confocal microscope system. The excitation/emission settings for detecting red and green fluorescence were 560/595 nm and 485/535 nm respectively.

*3.2.9 Quantification of mitochondrial fluorescence*

The mitochondrial membrane potential ($\Delta\Psi_m$) was measured by TMRE probe, which is a cell permeable, cationic, nontoxic, fluorescent dye that specifically stains live mitochondria in proportion to the membrane potential. Briefly, cells cultured in a specific temperature-controlled culture dish (MatTek, MA, USA) were incubated with TMRE (100 nM) in standard Tyrode solution containing (mM) NaCl 140, KCl 6, $MgCl_2$ 1, $CaCl_2$ 1, HEPES 5, and glucose 5.8 (pH 7.4) for 20 min. Cells were washed several times with fresh Tyrode solution. The dish was then observed by the Nikon A1 confocal microscope system.

*3.2.10 Statistical analysis*

For experimental data, all bar graphs represent the mean ± S.D. and all heatmaps represent the mean. To assess the statistical significance of differences between groups, paired or unpaired Student's t-test and one-way ANOVA test were utilized. All statistical analyses were carried out using R software v3.6.4 or GraphPad Prism v9.3. The following significance levels were used: * *p*<0.05; ** *p*<0.01.

# 4  Production of PSWCs

The pure water is stored in the water storage tank *1*, and purified water is pumped into the capillary array *3* by the water supply pump *2* (see Fig.2). The capillary is the anode connected to the positive electrode of the power supply *4*. The tips of the capillary array are aligned to the circular hole of the cathode *5*. The cathode is connected to the negative electrode of the power supply. The flow rate of purified water is 12 ml/min, the voltage of the power supply (homemade) is 10,000 volts. Water is electrosprayed from the Taylor cone, generating protonated small water clusters, which are collected by the water storage tank *6*.

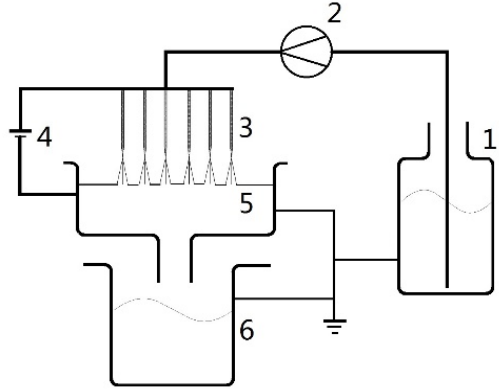

**Fig. 2** Schematic diagram of production equipment, where, 1. water storage tank; 2. water supply pump; 3 capillary array; 4. Power supply; 5. Cathode having circular holes corresponding to capillary array; 6. water storage tank.

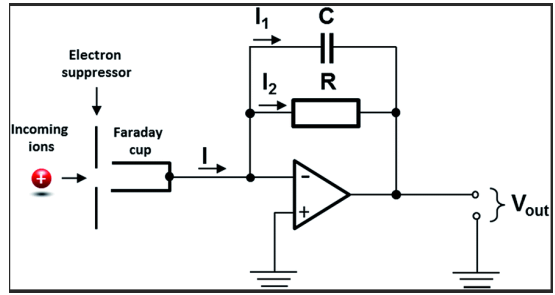

**Fig. 3** Electrostatic test principle diagram.

## 5  Measurement of the Charges of PSWCs

According to the electrostatic test principle diagram shown in Fig. 3, water droplets containing **PSWCs** with a hydrogen ion concentration of $10^{-4}$ mol/L or $10^{-3}$ mol/L, respectively in a volume of 10 microliters per drop are introduced into the Faraday cup with a quantitative dropper, and the charges measured are shown in Table 2. It can be seen that the water droplets do carry charges and the charge is proportional to the hydrogen ion concentration.

**Table 2** Measured water droplet charges from two different hydrogen ion concentrations.

|  | 1 | 2 | 3 | 4 | 5 | Mean value (nC) | Standard error |
|---|---|---|---|---|---|---|---|
| $10^{-4}$ mol/L | 0.1384 | 0.1429 | 0.1574 | 0.1484 | 0.1462 | 0.1466 | 0.0071 |
| $10^{-3}$ mol/L | 0.1914 | 0.1784 | 0.1884 | 0.1984 | 0.1884 | 0.1890 | 0.0072 |

## 6　Measurement of the Charges of PSWCs

An electrospray mass spectrometer was used to test the molecular constituents of the protonated small water clusters. In order to eliminate the influence of the laboratory environment atmosphere on the analysis results, an enclosed nano ESI source (CEESI ion source, Zhejiang Haochuang Biotech Co. Ltd., Hangzhou, China) with carrier gas flow was used to control the atmosphere around the spray tip. Once the voltage of electrospray ion source is too high, the molecular structure of charged small molecular clusters of hydronium ions will be destroyed. The spray voltage is set at 1.05x of the threshold voltage ($V_0$) of electrospray, and the voltage is adjustable. Supposing that the voltage of electrospray just occurred is the threshold voltage ($V_0$), and the experimental results show that the threshold voltage is 1.2 kV, so the spray voltage is set to 1.26 kV. Generally, the temperature of the heater capillary tube of the mass spectrometer is set at about 200 ℃. In order to eliminate the influence of high temperature on the molecular structure of protonated small water clusters, the temperature of the heater capillary tube is set at 30℃. The syringe pump is running at 2 μl/min. The obtained mass spectrum is shown in Fig. 4. It is clearly seen that the observed PSWCs are composed of water molecules with a formula $H^+(H_2O)_n$, (n = 2, 3, 4, 5), respectively, indicating the formation of the small water molecule clusters with protons.

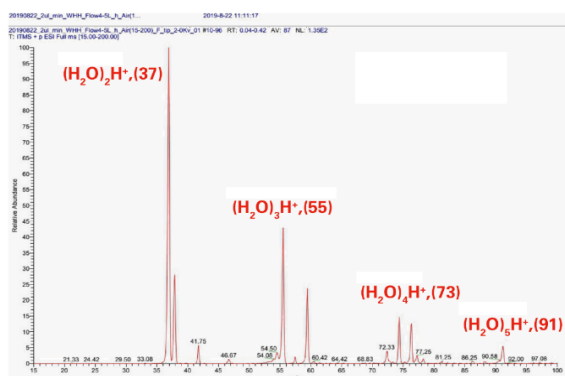

**Fig. 4** Mass spectrum of **PSWCs**.

It is clearly seen that the observed PSWCs are composed of water molecules with a formula $H^+(H_2O)_n$, (n = 2, 3, 4, 5), respectively, indicating the formation of the small water molecule clusters with a protons.

# 7．Experiment of Characteristics and Biological effects of PSWCs

## 7.1 Effects of PSWCs on the symptoms of DSS-induced colitis in mice

To investigate its therapeutic potential, we treated mice with protonated small water clusters with two $H^+$ concentrations ($10^{-3}$ and $10^{-4}$ mol/L) or Infliximab (as a positive control group) during DSS administration to induce colitis (Fig. 5 (A)). Mice treated with protonated small water clusters exhibited attenuated signs of colitis compared to the control group, evidenced by less weight loss, decreased disease activity index, longer colon length, and less severe histopathological (Fig. 5 (B-G)). Therefore, it is seen that **PSWCs** can effectively relieve the symptoms of DSS-induced colitis in mice.

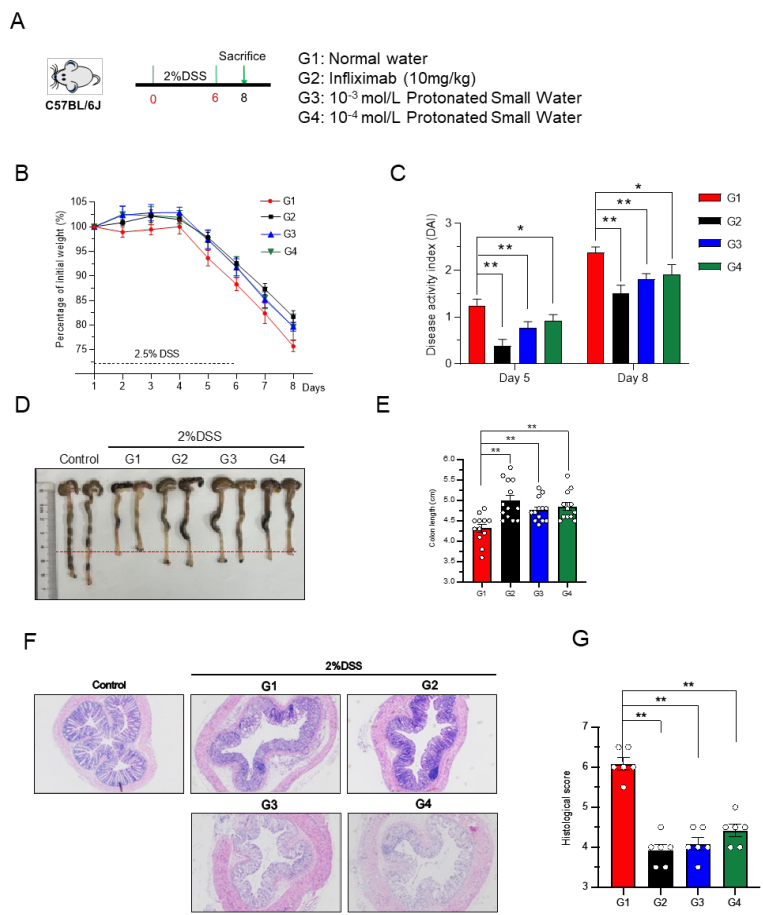

**Fig. 5** PSWCs can effectively relieve the symptoms of DSS-induced colitis in mice. (A) Scheme of Protonated Small Water treatment during the 2% DSS induced-colitis. G1 for the negative control; G2 Infliximab group for positive control; G3 and G4 for the experiment groups. (B-C) Body weight record (B) and disease activity index (C) of different groups mice during 2% DSS treatment, n=6. (D-G) Representative colon image (D), colon length (E), representative H&E staining image (F) and histological score of colonic section (G) from different group of mice on day 8 post 2% DSS administration, n=6. Data are presented as mean±SD; * $p<0.05$; ** $p<0.01$ by one-way ANOVA test.

*7.2 Effects of PSWCs on the expression level of inflammatory factors in colon after DSS-induced colitis*

It is clearly seen that the observed PSWCs are composed of water molecules with a formula $H^+(H_2O)_n$, (n = 2, 3, 4, 5), respectively, indicating the formation of the small water molecule clusters with protons.

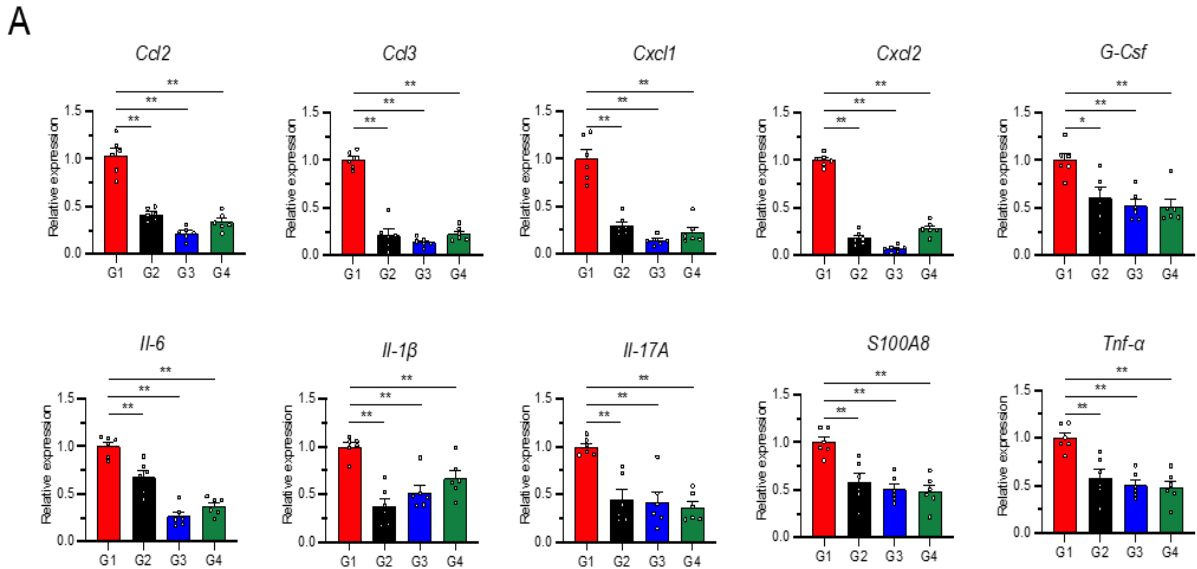

**Fig. 6** Protonated Small Water can effectively reduce the expression level of inflammatory factors in colon after DSS-induced colitis. (A) Quantitative mRNA expression of inflammatory chemokines and cytokines in colon from different groups mice on day 8 post 2% DSS treatment, n=6. Data are presented as mean ±S.D. * p<0.05; ** p<0.01 by one-way ANOVA test.

*7.3 Effects of PSWCs on ROS levels in the colon*

It is clearly seen that the observed PSWCs are composed of water molecules with a formula $H^+(H_2O)_n$, (n = 2, 3, 4, 5), respectively, indicating the formation of the small water molecule clusters with protons. To further explore how **PSWCs** affects colitis, we analyzed their effects on ROS levels, mitochondrial related gene expression, mitochondrial respiratory chain related protein levels, and mitochondrial membrane potential in intestinal epithelial tissues after DSS treatment. The result showed that compared with normal mice, DSS induction could significantly promote ROS levels in intestinal epithelial tissues. Positive drug Infliximab could inhibit the up-regulation of intestinal ROS induced by DSS, but had no significant effect. On the other hand, mice treated with **PSWCs** showed significantly inhibition of the DSS induced up-regulation of ROS in intestinal tissues (Fig.7(A-B)).

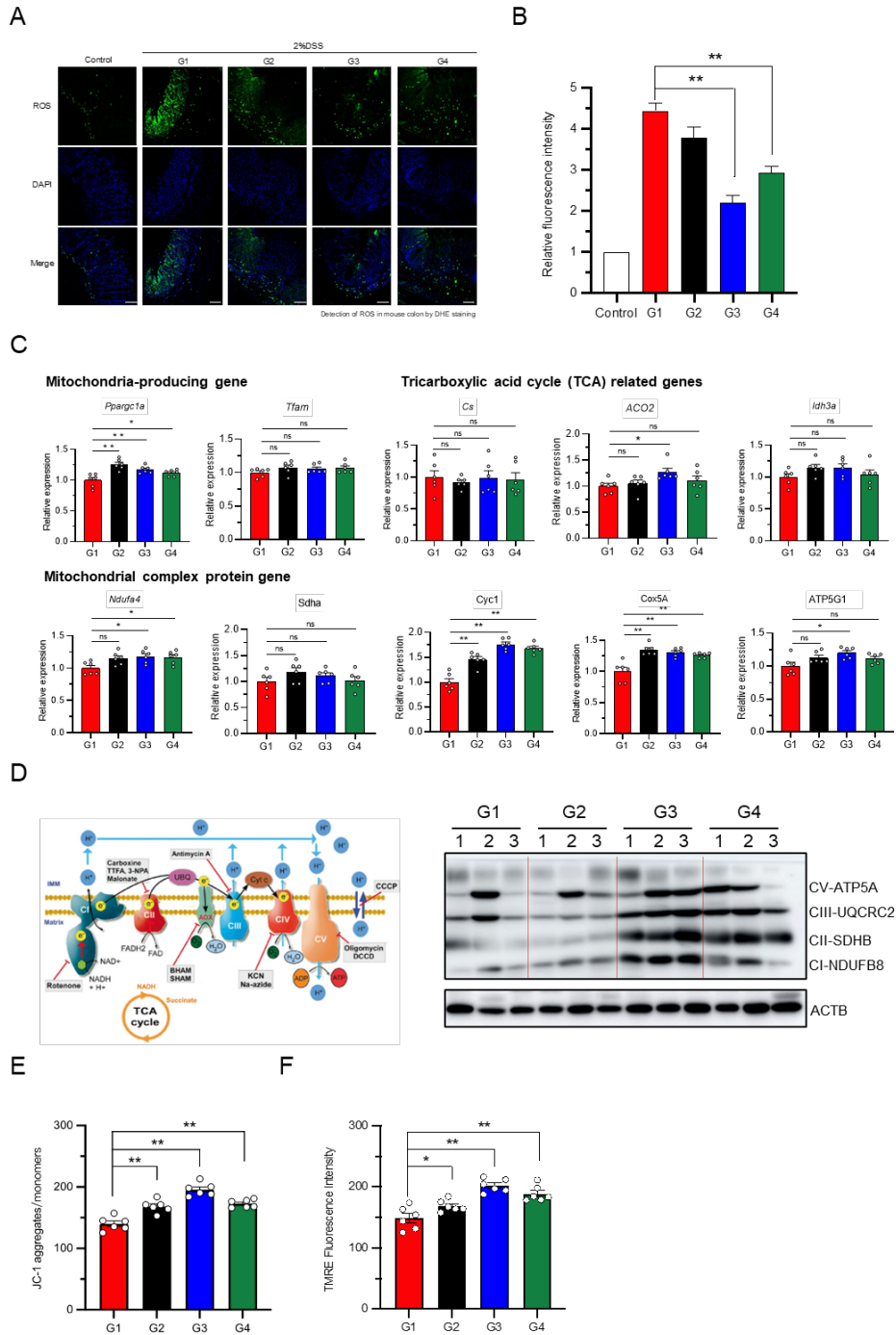

**Fig. 7** Protonated Small Water can effectively reduce ROS levels in the colon (A) ROS staining representation of mouse intestinal epithelium. (B) ROS level statistical results. (C) Expression levels of mitochondria-related genes in mouse colorectal tissues. (D) Schematic diagram and content level of mitochondrial respiratory chain related proteins by immunoblotting in mouse colorectal tissues. (E) Statistical results of JC-1 staining. (F) Statistical results of TMRE detection. Data are presented as mean±SD. * $p<0.05$; ** $p<0.01$ by one-way ANOVA test.

The strong link between ROS and mitochondria prompted us to investigate the effect of **PSWCs** on mitochondria. For this purpose, we used RT-qPCR to analyze the expression of mitochondrial related genes. The results showed that treating mice with **PSWCs** could affect the expression of mitochondrial genes in intestinal epithelial tissues. It significantly promoted the expression levels of mitochondria-producing genes *Ppargc1a*, mitochondrial complex protein genes *Ndufa4*, *Cyc1* and *Cox5a* (Fig. 6. (C)). Meanwhile, the levels of mitochondrial respiratory chain related proteins were analyzed by immunoblotting. The results showed that treating mice with **PSWCs** could significantly increase the levels of mitochondrial respiratory chain related proteins in intestinal epithelial tissues (Fig. (D)). Finally, we used membrane permeability JC-1 staining and fluorescent probe TMRE staining to detect the mitochondrial membrane potential in intestinal epithelium. The results showed that treating mice with **PSWCs** could significantly increase the level of mitochondrial membrane potential (Fig.s 7(E-F)).

Collectively, these results clearly reveal that treating mice with **PSWCs** can promote the formation and function of mitochondria, thereby down-regulate the ROS level in intestinal epithelial tissues and maintain their integrity.

*7.4 Effects of PSWCs on the expression level of inflammatory factors in liver after DSS treatment*

The intestine and liver are vital immune organs, and they interact and communicate through immune cells, cytokines, and other factors. Immune abnormalities triggered by colitis may disrupt the balance of immune system, thus affecting the liver's immune response and inflammatory regulation. We analyzed the effect of **PSWCs** on the liver. The results showed no significant differences in size and weight of liver (Fig. 8(A-B)). Moreover, there were no significant differences in morphology, lipid droplet formation and the levels of triglycerides in mice liver. (Fig. 8(C-F)).

We then analyzed the inflammatory factors including inflammatory chemokines and cytokines. RT-qPCR result revealed that treating mice with **PSWCs** decreased the expression of inflammatory factors in the liver (Fig. 8(E)). Together, these results indicate treating mice with **PSWCs** could also suppress the inflammatory levels in the liver, but did not affect other indicators.

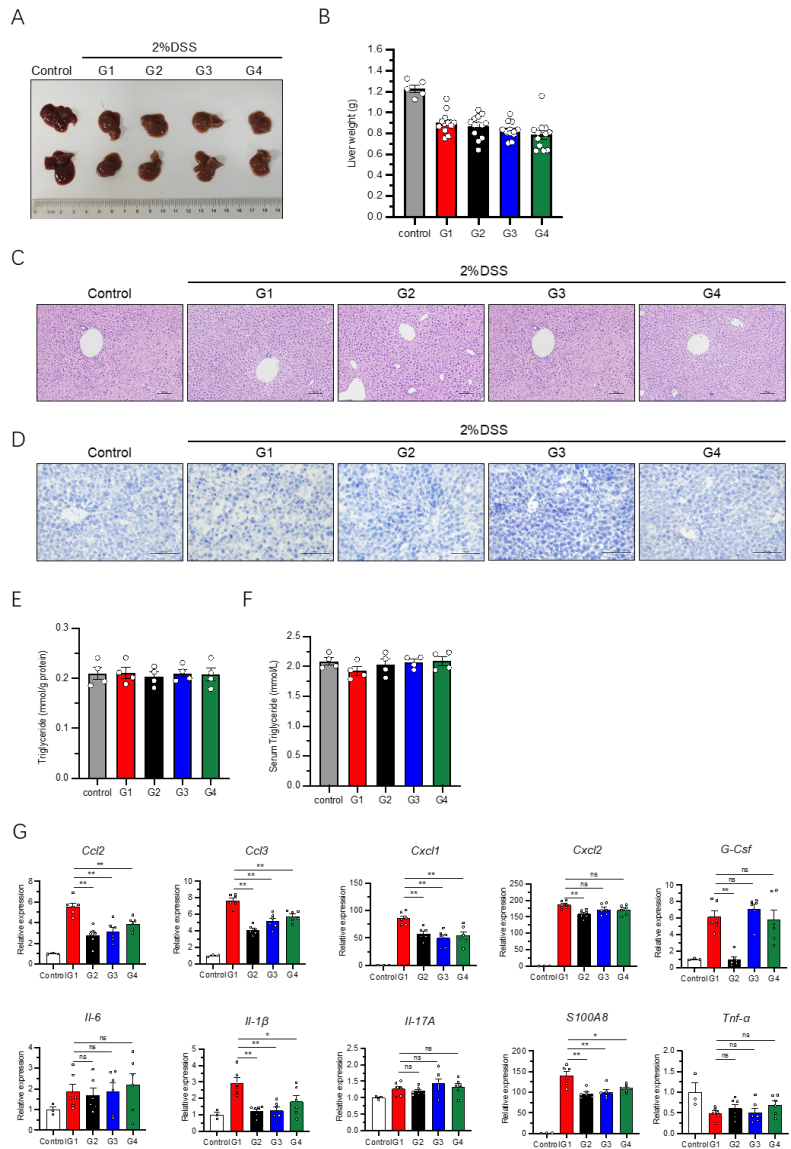

**Fig. 8** Protonated Small Water can effectively reduce the expression level of inflammatory factors in liver after DSS treatment.

### 7.5 Effects of PSWCs on ROS levels in the liver

Furthermore, we evaluated the effects of **PSWCs** on ROS levels in the liver after DSS treatment, including the levels of ROS in liver, the expression of mitochondrial-related genes, the levels of mitochondrial respiratory chain-related proteins, and the impact on mitochondrial membrane potential. The results showed that compared with normal mice, DSS induction can significantly promote ROS levels in liver tissues. Positive drug Infliximab showed down-regulation of liver ROS induced by DSS, but has no significant effect. On the other hand, treating mice with protonated small water clusters significantly inhibited the DSS induced up-regulation of ROS in liver tissues (Fig.9(A-B)).

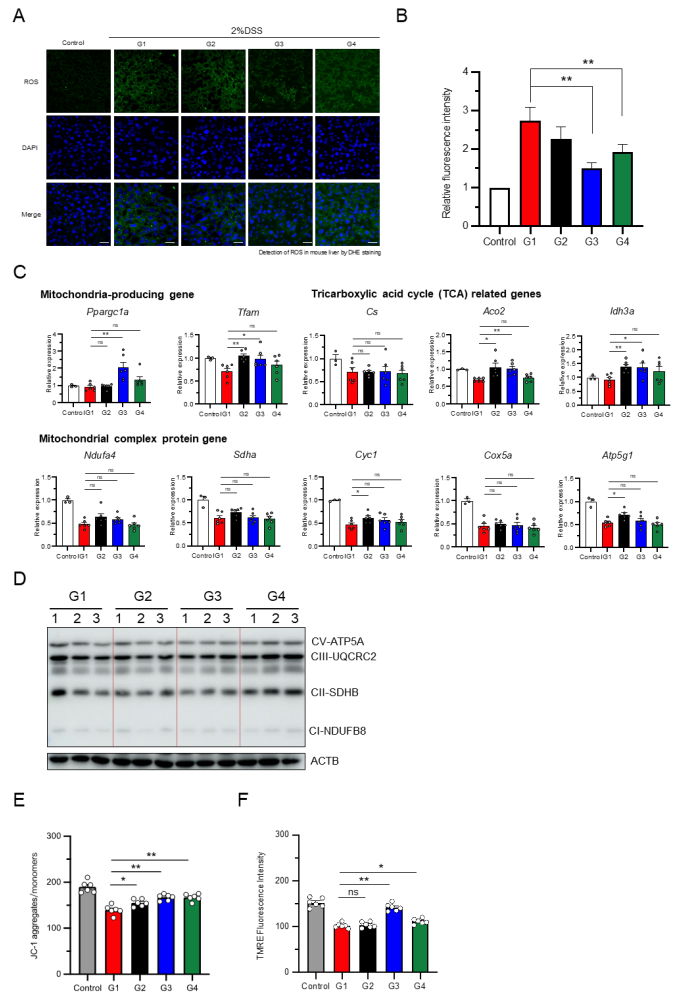

**Fig. 9** Protonated Small Water can effectively reduce ROS levels in the liver (A) Representative ROS staining of mouse liver. (B) ROS level statistical results. (C) Expression levels of mitochondria-related genes in mouse liver tissues. (D) Representative immunoblotting pictures of mitochondrial respiratory chain related proteins in mouse liver tissues. (E) Statistical results of JC-1 staining. (F) Statistical results of TMRE detection. Data are presented as mean±SD. * $p<0.05$; ** $p<0.01$ by one-way ANOVA test.

Meanwhile, we used RT-qPCR to analyze the expression of mitochondrial related genes. The results showed that treating mice with **PSWCs** at $10^{-3}$ mol/L hydrogen ion concentration significantly influenced the expression levels of mitochondrial biogenesis genes *Ppargc1a* and *Tfam*, while treating mice at $10^{-4}$ mol/L hydrogen ion concentration showed certain trend but no significance (Fig. 9(C)). Besides, the levels of mitochondrial respiratory chain related proteins were analyzed by immunoblotting. The results showed that treating mice with **PSWCs** did not affect the levels of mitochondrial respiratory chain related proteins in liver tissues (Fig. 9(D)). Finally, we used membrane permeability JC-1 staining and fluorescent probe TMRE staining to detect the mitochondrial membrane potential in liver. The results showed that treating mice with **PSWCs** could significantly increase the level of mitochondrial membrane potential (Fig. 9(E-F)). Collectively, these results clearly revealed that treating mice with **PSWCs** can promote the formation and function of mitochondria, downregulating the ROS level in liver tissue and reducing liver inflammation, thereby protecting liver functions.

# 8   Conclusions

In summary, we have reported the production, characteristics and the biological effects of the **PSWCs**. We first generate the **PSWCs** by electrospray mechanism, which is proved to be charged (with positive charge). We found that the **PSWCs** are composed of 2, 3, 4, 5 water molecules plus a proton, respectively, which has long-term stability without any proton concentration change. Furthermore, we perform the biological function tests of **PSWCs,** the results show many very important biological functions, such as **PSWCs** can reduce the radicals (such as reactive oxygen species), and improve the cell functions, etc.. As typical examples, **PSWCs** can significantly inhibit the level of ROS in intestinal epithelium and liver tissues induced by DSS, promote the expression levels of mitochondria-related genes and proteins, and the level of mitochondrial membrane potential. It should be noted that, during DSS induced mouse colitis, treating with **PSWCs** improved the symptoms of colitis. By examining body weight, disease activity index, colorectal length, and intestinal inflammation related indicators, we can conclude that **PSWCs** have a potential therapeutic effect on colitis. This indicates that, **PSWCs** could significantly inhibit the level of ROS in intestinal epithelium and liver tissues induced by DSS, and promote the expression levels of mitochondria-related genes and proteins, and the level of mitochondrial membrane potential.

**Disclosures**

No conflicts of interest, financial or otherwise, are declared by the authors.

**Acknowledgments**

We thank Jun Sun and Yifan Liu (Zhejiang University School of Medicine) for technical assistance in animal experiments.

*References*

1. Nageswara R. Madamanchi, Marschall S. Runge, "Mitochondrial Dysfunction in Atherosclerosis," *Circ. Res.*, **100**(4), 460-473 (2007).
2. Nadja B. Cech and Christie G. Enke, "Practical implications of some recent studies in electrospray ionization fundamentals" MASS SPECTROM REV., **20**, 362-387(2001).
3. Ohsawa, I.; Ishikawa, M.; Takahashi, K.; Watanabe, M.; Nishimaki, K.; Yamagata, K.; Katsura, K.; Katayama, Y.; Asoh, S.; Ohta, S. *"*Hydrogen acts as a therapeutic antioxidant by selectively reducing cytotoxic oxygen radicals," *Nat. Med.* **13** (6), 688−694 (2007).

**Author** is the Chairman of Hangzhou Shanshangshui Technology Co., Ltd. in China. He received his Bachelor's degree in Engineering in 1982 and his Master's degree in engineering in 1989 from Zhejiang University in China. He is the inventor of Bruker's CaptiveSpray and holds more than 20 patents in the United States and China.

**Caption List**

**Fig. 1** Schematic diagram of (**PSWCs**).

**Fig. 2** Schematic diagram of production equipment, where, 1. water storage tank ; 2. water supply pump; 3 capillary array; 4. the positive electrode of the power supply; 5. circular hole of the cathode;6. the water storage tank.

**Fig. 3** Electrostatic test principle diagram.

**Fig. 4** Mass spectrum of **PSWCs**.

**Fig. 5** PSWCs can effectively relieve the symptoms of DSS-induced colitis in mice (A) Scheme of Protonated Small Water treatment during the 2% DSS induced-colitis. G1 for the negative control; G2 Infliximab group for positive control; G3 and G4 for the experiment groups. (B-C) Body weight record (B) and disease activity index (C) of different groups mice during 2% DSS treatment, n=6.

(D-G) Representative colon image (D), colon length (E), representative H&E staining image (F) and histological score of colonic section (G) from different group of mice on day 8 post 2% DSS administration, n=6. Data are presented as mean ±SD; * *p*<0.05; ** *p*<0.01 by one-way ANOVA test.

**Fig. 6** Protonated Small Water can effectively reduce the expression level of inflammatory factors in colon after DSS-induced colitis (A) Quantitative mRNA expression of inflammatory chemokines and cytokines in colon from different groups mice on day 8 post 2% DSS treatment, n=6. Data are presented as mean±SD. * p<0.05; ** p<0.01 by one-way ANOVA test.

**Fig. 7** Protonated Small Water can effectively reduce ROS levels in the colon (A) ROS staining representation of mouse intestinal epithelium. (B) ROS level statistical results. (C) Expression levels of mitochondria-related genes in mouse colorectal tissues. (D) Schematic diagram and content level of mitochondrial respiratory chain related proteins by immunoblotting in mouse colorectal tissues. (E) Statistical results of JC-1 staining. (F) Statistical results of TMRE detection. Data are presented as mean ±SD. * *p*<0.05; ** *p*<0.01 by one-way ANOVA test.

**Fig. 8** Protonated Small Water can effectively reduce the expression level of inflammatory factors in liver after DSS treatment.

**Fig. 9** Protonated Small Water can effectively reduce ROS levels in the liver (A) Representative ROS staining of mouse liver. (B) ROS level statistical results. (C) Expression levels of mitochondria-related genes in mouse liver tissues. (D) Representative immunoblotting pictures of mitochondrial respiratory chain related proteins in mouse liver tissues. (E) Statistical results of JC-1 staining. (F) Statistical results of TMRE detection. Data are presented as mean±SD. * *p*<0.05; ** *p*<0.01 by one-way ANOVA test.

**Table 1** RT-qPCR Primers.

**Table 2** Measured water droplet charges from two different hydrogen ion concentrations.